\documentclass{ametsocV6.1}

\title{Spatial Transferability of Machine Learning Downscaling Under Varying Surface and Orographic Conditions}

\authors{Lewis Blunn,\aff{a}\correspondingauthor{Lewis Blunn, lewis.blunn@metoffice.gov.uk} 
Alfie Bull Green,\aff{a} 
Tomas Wetherell,\aff{a} 
 and Ben Booth\aff{a} 
}

\affiliation{\aff{a}{Met Office, Exeter, United Kingdom}\\
}

\abstract{
High-resolution weather information is needed to understand local variations in hazards and impacts, but producing
it with conventional numerical models is computationally expensive. Machine-learning downscaling offers a faster alternative by learning to reconstruct fine-scale weather patterns from coarser model output. However, potential training datasets are available for only limited spatial domains, especially as we move to higher-resolution, urban-scale predictions, and it remains unclear whether the relationships they learn can be transferred reliably to locations with different land-surface and topographic characteristics. Here we show that a deep-learning model trained on only a few
small areas can downscale near-surface air temperature from 2.2~km to 300~m resolution across previously unseen locations within the Paris region. Fine-scale temperature patterns associated with urban surfaces, water and terrain generalise rapidly, while overall performance stabilises after training on approximately five spatial patches and improves on the 2.2~km model by 5.6\%. Successful transfer requires explicit high-resolution information about the land surface and the differences between coarse- and fine-resolution topography. The model generalises well to slopes outside its training range,
but training solely on low-lying areas produces a warm bias over higher terrain, showing that representative elevation regimes remain necessary. These results identify the physical information and modest spatial diversity needed for transferable machine-learning downscaling, supporting its use to extend high-resolution weather information across regions where local training simulations are limited.
}

\begin{document}

\maketitle

\statement
   Machine learning downscaling models offer a cost-effective path to fine-scale local weather forecasts, but training them typically requires detailed data unavailable globally. This study demonstrates that deep learning models can downscale near-surface air temperature (from 2.2~km to 300~m) and transfer this skill to unseen regions after training on as few as five localized patches. Achieving robust transferability relies on incorporating high-resolution static surface descriptors, residual surface descriptors capturing grid resolution differences, and representative elevation ranges during training. These findings highlight the potential for expanding high-resolution ML weather and climate predictions and projections across large, data-sparse domains.

\section{Introduction}\label{sec:intro}

Recent years have seen rapid advances in machine learning (ML) applications for weather forecasting \citep{rabier2026weather}. In particular, machine-learning weather prediction (MLWP) has demonstrated considerable promise at both global and regional scales, with performance that is comparable to, and in many cases surpassing conventional numerical weather prediction (NWP) systems on certain key metrics \citep{lam2023learning,ben2024rise,abdi2026hrrrcast,nipen2026regional}.

Alongside these developments, machine learning downscaling (MLD) has emerged as an active research area in weather and climate science \citep{harris2022generative,rampal2024enhancing,mardani2025residual,addison2026machine,dumont2026downscaling}. Rather than predicting future atmospheric states, MLD transforms coarse forecasts, projections, or analyses into higher-resolution fields by adding fine-scale spatial detail and correcting systematic biases. Inputs from conventional or ML-based simulations are supplemented with surface descriptors to generate high-resolution predictions at a fraction of the computational cost of dynamical downscaling -- critical for localized hazard prediction, urban weather services, and climate applications. MLD models can emulate high-resolution NWP simulations \citep{doury2023regional} or integrate observations directly to correct systematic dynamical errors \citep{blunn2024machine,Li2026}.

The urban-scale modelling project at the Met Office aims to exploit MLD to reproduce the benefits of hectometre-scale NWP across large regional domains at reduced computational cost. Because running high-resolution NWP continuously over broad areas is computationally prohibitive, ML models must be trained on localized domains and transferred zero-shot to unobserved regions. Here, zero-shot is used to describe application of a model to spatial regions that were withheld entirely from training. This introduces the challenge of spatial transferability: the extent to which relationships learned in one location generalise to regions with differing surface, orographic, and climatic characteristics.

Prior research on spatial transferability has largely focused on precipitation and near-surface temperature over multiple geographically and climatologically distinct domains. Benchmarking by \citet{harder2025rainshift,harder2025global} shows that generative precipitation downscaling improves upon spatial interpolation in unseen regions, yet generalisation degrades significantly when transferring across distinct orographic or climatic regimes. Similarly, \citet{blasone2024deep} observed performance drop-offs as test locations moved further from the training domain, though models retained skill in capturing heavy precipitation driven by large-scale dynamics. For near-surface air temperature, \citet{benitez2025transferability} demonstrated that U-Net downscaling generalises well across moderate terrain, but incurs localized error inflation over high-relief features outside the training distribution, such as the Rocky Mountains. Collectively, these studies indicate that spatial transferability is achievable, but remains highly sensitive to orographic and land-surface shifts.

To our knowledge, no previous study has systematically isolated the influence of surface characteristics and orography on spatial transferability within a geographically and climatologically similar domain, thereby removing the confounding effects of broad climatic differences.

In this study, we investigate the zero-shot spatial transferability of MLD for 1.5~m air temperature (hereafter, temperature) using a unique dataset of 300~m ensemble simulations over Paris, produced during the Paris 2024 Olympics Research Demonstration Project \citep{masson2026}. The domain is partitioned into a $5 \times 5$ patch grid \citep{benitez2025transferability,harder2025rainshift,harder2025global}. By training models on varying patch subsets and evaluating on held-out patches, we address the following research questions:
\begin{enumerate}
\item How much spatial diversity is required in the training dataset to achieve robust zero-shot transferability to previously unseen locations?
\item To what extent does model performance degrade when applied to locations whose surface and orographic characteristics lie outside the training distribution?
\item Can high-resolution residual surface descriptors improve transferability to locations with contrasting orographic characteristics and steep-terrain conditions?
\end{enumerate}

Addressing these questions provides practical insights for deploying MLD where local training data are sparse or unavailable, supporting initiatives like the World Meteorological Organization's Early Warnings for All \citep{WMO2022,UrbanPREDICT2025} by expanding access to high-resolution weather information globally.

\section{Experimental Design and Methodology}\label{sec:expdesign}

\subsection{Conventional Model Simulations}\label{sec:convmodel}

During the summer of 2024, the Met Office conducted real-time 300~m ensemble forecasts over Paris \citep{masson2026} as part of the Paris 2024 Olympics Research Demonstration Project. The MOGREPS-G ensemble forecasts \citep{bowler2008mogreps} were initialized using 4DVAR analyses \citep{milan2020hourly} at 18:00~UTC daily, with 3-hour lead-time output initialising a nested 2.2~km dynamical downscaler, which in turn drove an inner nested 300~m model. Both nested domains were centered on Paris and featured variable-resolution boundaries to enable efficient boundary-layer spin-up before reaching their fixed-resolution core regions: $500 \times 499$ grid points ($1,100\text{ km} \times 1,100\text{ km}$) for the 2.2~km model and $450 \times 449$ grid points ($150\text{ km} \times 150\text{ km}$) for the 300~m model. The nested models ran Unified Model version 13.0 \citep{davies2005new,wood2014inherently} with Regional Atmosphere and Land configuration 3.2 \citep{bush2024third}, coupled to the JULES land surface model \citep{best11,clark11} and the MORUSES urban scheme \citep{porson10}.

Forecasts were initialized at 21 UTC daily from 16 June to 9 September as 18-member ensembles out to 36-hour lead times. This study analyzes forecasts from 14 August to 10 September 2024 to ensure the land cover dataset used in the simulations is consistent, following an update to the WorldCover urban fraction dataset on 12 August \citep{zanaga2022esa}. Missing initializations on 19 August and 2 September reduce sample counts on those and subsequent two days. Non-urban land cover tiles are derived from ESA CCI land cover \citep{esa2017}, and topographies originate from the 3-arc-second SRTM dataset \citep{NASA2013}.

\subsection{Problem Formulation and Input Predictors}\label{sec:predictors}

The MLD framework emulates the 300~m reference temperature using 300~m static surface descriptors (and their differences from the corresponding 2.2~km static surface descriptors) and 2.2~km atmospheric fields as inputs. The 300~m and 2.2~km variables are obtained from their respective resolution models, rather than by coarse-graining the 300~m model fields to 2.2~km, as done in some approaches. This choice is motivated by the dual objective of the MLD model: it must not only add fine-scale spatial detail but also learn and correct systematic biases between the 2.2~km atmospheric input fields and the 300~m reference.

The output is the residual temperature, defined as the 300~m reference temperature minus the bilinearly interpolated 2.2~km atmospheric input temperature. During inference, absolute high-resolution temperatures are reconstructed by adding the interpolated 2.2~km atmospheric input field to the predicted residual field.

The input fields comprise nine prognostic atmospheric variables from the 2.2~km atmospheric input fields and six 300~m static surface descriptors (Table \ref{tab:predictors}). The 2.2~km prognostic variables were selected based on expert domain knowledge regarding the governing near-surface thermodynamics and boundary-layer dynamics. The 300~m static descriptors (orography, urban fraction, and inland water fraction) capture localized spatial heterogeneity. Surface descriptor residuals (300~m minus 2.2~km representations of orography, urban fraction, and inland water fraction) are included to directly provide the model with resolution discrepancies. Throughout, slope is defined as the modulus of the orography gradient. All input and reference fields are standardized using the mean and standard deviation calculated strictly from the training partition.

\begin{table}[ht]
\caption{Summary of predictor and reference variables used in the MLD framework.}\label{tab:predictors}
\begin{center}
\begin{tabular}{lll}
\hline
\textbf{Variable Category} & \textbf{Feature Description} & \textbf{Units / Range} \\ \hline
\textbf{2.2~km Atmospheric Predictors} & 1.5~m Temperature & $^\circ$C \\
& 10~m $x$ Wind Component & m s$^{-1}$ \\
& 10~m $y$ Wind Component & m s$^{-1}$ \\
& Surface Downwelling Shortwave Radiative Flux & W m$^{-2}$ \\
& Surface Downwelling Longwave Radiative Flux & W m$^{-2}$ \\
& Surface Upward Sensible Heat Flux & W m$^{-2}$ \\
& Surface Upward Latent Heat Flux & W m$^{-2}$ \\
& Soil Moisture (top 10~cm) & m$^3$ m$^{-3}$ \\
& Stratiform Rainfall Flux & mm h$^{-1}$ \\
& Cloud Area Fraction (maximum random overlap) & \% \\ \hline
\textbf{300~m Static Surface Predictors} & Orography & m \\
& Orography Residual (300~m $-$ 2.2~km) & m \\
& Urban Fraction & 0 to 1 \\
& Urban Fraction Residual (300~m $-$ 2.2~km) & $-$1 to 1 \\
& Inland Water Fraction & 0 to 1 \\
& Inland Water Fraction Residual (300~m $-$ 2.2~km) & $-$1 to 1 \\ \hline
\textbf{300~m Reference} & 1.5~m Residual Temperature ($\Delta T_{1.5\text{m}}$) & $^\circ$C \\ \hline
\end{tabular}
\end{center}
\end{table}

\subsection{Data Partitioning and Experimental Setups}

\subsubsection{Temporal Splitting}

The dataset (14 August to 10 September 2024) utilizes a chronological train-validation-test split (Table \ref{tab:datasplit}) starting after the 12 August land cover update to maintain static descriptor consistency. The test window was selected to include the domain-mean temperature maximum and minimum, providing a stricter evaluation under both warm and cool extremes than would have been the case if the end of the dataset had been used for the test period. Because this places test data mid-dataset, we inserted two validation buffers (29--30 August and 6--7 September), giving a 48-h separation on each side between train and test. This prevents information leakage between the train and test periods. Accounting for missing initialization runs on 19 August and 2 September, the total training, validation, and testing sample sizes are 10,674, 2,520, and 3,150 frames, respectively.

\begin{table}[ht]
\caption{Temporal partition and sample sizes of the dataset.}\label{tab:datasplit}
\begin{center}
\begin{tabular}{llc}
\hline
\textbf{Dataset Split} & \textbf{Calendar Dates (2024)} & \textbf{Total Samples ($N$)} \\ \hline
\textbf{Training} & 14 Aug -- 28 Aug (15 days) \& 08 Sept -- 10 Sept (3 days)$^\dagger$ & 10,674 \\
\textbf{Validation} & 29 Aug -- 30 Aug (2 days) \& 06 Sept -- 07 Sept (2 days) & 2,520 \\
\textbf{Testing} & 31 Aug -- 05 Sept (6 days)$^\ddagger$ & 3,150 \\ \hline
\multicolumn{3}{l}{\small $^\dagger$ Excludes missing 19 August initialization. $^\ddagger$ Excludes missing 02 September initialization.}
\end{tabular}
\end{center}
\end{table}

\subsubsection{Spatial Partitioning and Patch Layout} \label{sec:patches}

\begin{figure}[ht]
  \centerline{\includegraphics[width=1.0\textwidth]{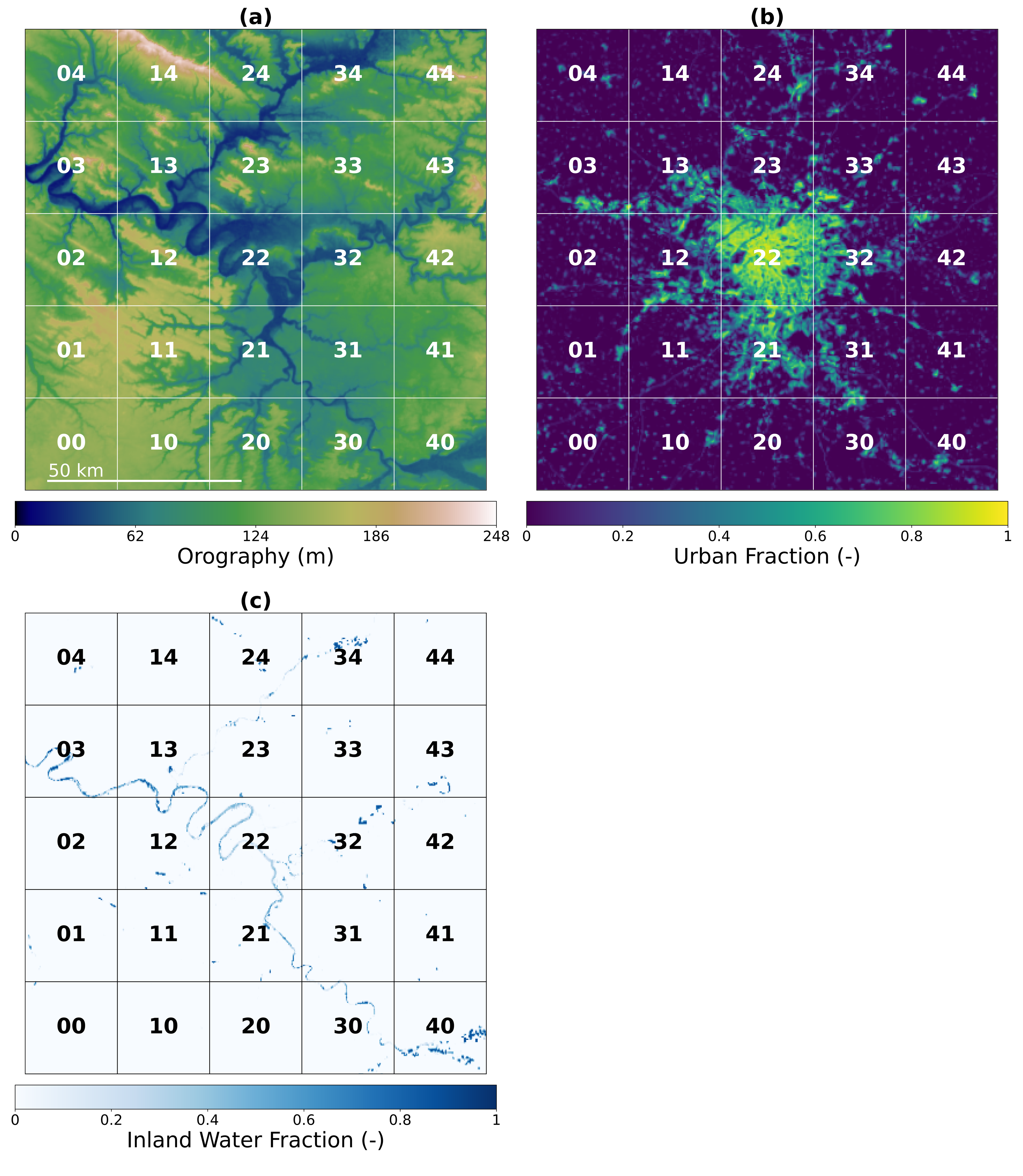}}
  \caption{(a) Orography, (b) urban fraction, and (c) inland water fraction across the 25-patch layout. Indexing follows $XY$ coordinates ($X, Y \in [0, 4]$).}\label{f1}
\end{figure}

The 2.2~km input fields are bilinearly interpolated onto the 300~m grid. The central $400 \times 400$ grid point region is partitioned into a $5 \times 5$ array of 25 spatial patches ($80 \times 80$ grid points each; Fig.~\ref{f1}). Patches are indexed via a two-digit $XY$ coordinate system, where $X$ increases left-to-right (0 to 4) and $Y$ increases bottom-to-top (0 to 4).

We track surface feature similarity between training patches and the test set using two coverage metrics:
\begin{itemize}
\item \textbf{Distribution Coverage Score (DCS):} computed by comparing normalized training and test histograms on shared bin edges defined over the union of training and test value ranges. Specifically, we compute the mean absolute bin-wise difference and rescale it to a similarity score, where 1 indicates identical distributions and 0 indicates maximal dissimilarity.
\[
\text{DCS} = 1 - \frac{1}{2}\sum_{b=1}^{B} \left|p_b^{\mathrm{train}} - p_b^{\mathrm{test}}\right|,
\]
where $p_b^{\mathrm{train}}$ and $p_b^{\mathrm{test}}$ are normalized frequencies in bin $b$ over $B$ shared bins.
\item \textbf{Range Coverage Ratio (RCR):} defined as the ratio of training-patch and test-set max-min ranges,
\[
\text{RCR} = \frac{\max(x_{\mathrm{train}})-\min(x_{\mathrm{train}})}{\max(x_{\mathrm{test}})-\min(x_{\mathrm{test}})}.
\]
Values exceeding 1 indicate that the training range encompasses the test range.
\end{itemize}

\subsubsection{Experimental Framework} \label{sec:expframework}

Three distinct controlled experiments are executed to target the core research objectives:
\begin{itemize}
    \item \textbf{Experiment 1: Scaling Spatial Training Diversity.} This experiment quantifies model sensitivity to the geographic diversity of the training data. The out-of-sample evaluation is performed over five fixed benchmark patches selected for their distinct environmental properties: patch 11 (semi-sparse urban, high orography), patch 14 (rural/sparse urban, very high orography), patch 22 (very dense urban intersected with river), patch 41 (rural/sparse urban, low orography, very low slope) and 43 (rural/sparse urban, low orography, high slope with narrow river channels). The training set expands progressively from 1 patch up to 20 patches using random selection, concluding with inclusion of the test patches in training. For each number of training patches, 10 randomized training realizations are performed, each using a different randomly selected combination of patches, to ensure results are robust against any single spatial configuration. Each realization also has different model weight initialization to account for stochasticity in training.
    \item \textbf{Experiment 2: Transferability Across Orographic-Complexity Shift.} This experiment isolates the impact of orographic complexity using the $99^{\text{th}}$ percentile of slope, computed across all grid points in each patch. The 25 patches are ranked by the $99^{\text{th}}$ percentile as a proxy for localized slope complexity. P99 was chosen to emphasize the upper tail rather than focussing on a single point extreme. An initial model is trained on the three flattest patches and evaluated on the three most orographically complex patches, representing a targeted train--test distribution shift along this terrain-complexity axis. A secondary configuration replaces one flat training patch with the steepest patch in the domain, reducing the terrain-complexity shift. For each configuration, three repeats are performed with different model weight initializations.
    \item \textbf{Experiment 3: Transferability Across Elevation Shift.} To evaluate model sensitivity to elevation differences, the 25 patches are ranked according to their mean orography. A baseline model is trained using only the three lowest-lying patches and tested on three high-elevation patches that lie beyond the training elevation range (withholding the absolute highest patch). To determine if spatial transferability constraints are alleviated by partial distribution exposure, a second model is trained by swapping out one low-elevation patch for the absolute highest-elevation patch, thereby reducing the train--test elevation shift. For each configuration, three repeats are performed with different model weight initializations.
\end{itemize}

To isolate structural surface diversity from training volume sample size effects, absolute training sample count is held constant ($N = 10,674$) across scaling experiments by randomly sub-sampling frames across times and ensemble members within the active training patches.

\subsection{Model Architecture}\label{sec:architecture}

To perform the spatial downscaling, we employ a deep learning approach based on the U-Net architecture originally developed by \citet{doury2023regional} for regional climate downscaling. We did not perform architecture search or exhaustive hyperparameter tuning, because the aim here is to isolate spatial transferability behavior using a peer-reviewed, well-performing baseline architecture rather than to optimize model design. The network features a symmetric encoder-decoder structure with four hierarchical levels and skip connections linking corresponding levels to preserve high-resolution spatial features. The encoding path begins with $64$ initial channels, with the channel depth doubling at each successive downsampling step.

All convolutional layers across both the encoder and decoder paths utilize a $3 \times 3$ kernel. To mitigate boundary-effect artifacts, reflection padding is applied. The final output layer consists of a $1 \times 1$ convolution that maps the latent features back to a single-channel residual temperature field.

We introduce two key structural adaptations to the baseline architecture by \citet{doury2023regional} to optimize it for our high-resolution meteorological data:
\begin{itemize}
    \item Removal of Terminal Up-sampling Blocks: Because the low-resolution 2.2~km predictor fields are bilinearly interpolated onto the 300~m reference grid prior to model input, the terminal up-sampling layers from the original architecture are redundant and thus omitted.
    \item Bilinear Decoder Interpolation: Within the decoding path, standard transposed convolutions are replaced with bilinear interpolation blocks. This architectural modification successfully eliminates checkerboard artifacts and grid-like distortions commonly associated with learned up-sampling layers.
\end{itemize}

\subsection{Model Training and Optimization}\label{sec:training}

The network is trained using a pointwise mean absolute error (MAE) objective, which optimizes for the conditional median of the target spatial field. Pointwise regression loss functions are known to smooth localized extremes, a problem that becomes particularly pronounced in multi-step autoregressive rollouts where spatial smoothing compounds over time. Here, downscaling operates as a direct single-step mapping, avoiding such compounding errors. Furthermore, sub-grid scale temperature heterogeneity below the driving model resolution is predominantly driven by persistent land-surface features (e.g., elevation, land cover) rather than fine-scale transient turbulent dynamics. While generative frameworks (such as diffusion models) can better represent fine-scale stochastic variability, a deterministic pointwise approach effectively captures the spatially persistent, surface-forced temperature signal. Crucially, because our primary objective is to evaluate spatial transferability, specifically the ability of the network to learn surface-driven physical relationships and generalize them to unseen locations, a deterministic pointwise loss function provides a robust and transparent baseline. 

Optimization is performed using the Adam optimizer with a batch size of $64$ and training was conducted for 20 epochs. After 8 epochs early stopping was used to save the model at the epoch with the lowest validation loss. To accelerate convergence and prevent overfitting, we employ a OneCycle learning rate schedule \citep{smith2019super} with a maximum learning rate of $5 \times 10^{-4}$.

\subsection{Evaluation Metrics}\label{sec:metrics}

We use the MAE between the reconstructed predictions ($\hat{y}_i$) and the 300~m reference temperature fields ($y_i$) as the primary metric for model performance:
\begin{equation}
\text{MAE} = \frac{1}{N} \sum_{i=1}^{N} |y_i - \hat{y}_i|
\end{equation}
where $N$ denotes the total number of evaluation grid points across the test patches.

For Fig.~\ref{fig:exp1_scaling}d, we also report a single mean temperature-offset value at each training-patch count, formed by first averaging the prediction fields over the full held-out test set for each of the 10 randomized training realizations and then aggregating those realization-specific offsets. This is therefore an aggregate diagnostic across the whole test set, not a pointwise or timestep-wise MAE. The signed mean difference is obtained by averaging the signed realization-specific offsets, so it shows the direction of the mean offset, that is, whether the model is systematically warm or cold relative to the reference field. The mean absolute difference is obtained by averaging the absolute realization-specific offsets, so it measures the typical size of the offset even when warm and cold biases cancel.

We also low-pass and high-pass filter the reference and predictions using a Fast Fourier Transform (FFT) to decompose the fields into coarse-scale and fine-scale components, with the cutoff at 2.2~km, which is the grid length of the atmospheric predictors. On the FFT we use reflect padding to mitigate edge effects. The coarse-scale component represents spatial structure at scales larger than the driving 2.2~km model resolution, whereas the fine-scale component isolates spatial detail at scales finer than 2.2~km. We inverse transform the bandpassed fields and compute the MAE for the coarse-scale and fine-scale components separately.

\section{Results and Discussion}\label{sec:results}

\subsection{Experiment 1: Scaling Spatial Training Diversity}\label{sec:exp1_results}

In this experiment, we investigate the influence of increasing the spatial diversity of the training data on the spatial transferability of the model. We train the network on an increasing number of spatial patches, scaling progressively from 1 up to 20 patches, before finally including the held-out test patches directly in the training set to establish an in-sample benchmark.

\begin{figure}[t]
  \centerline{\includegraphics[width=1.0\textwidth]{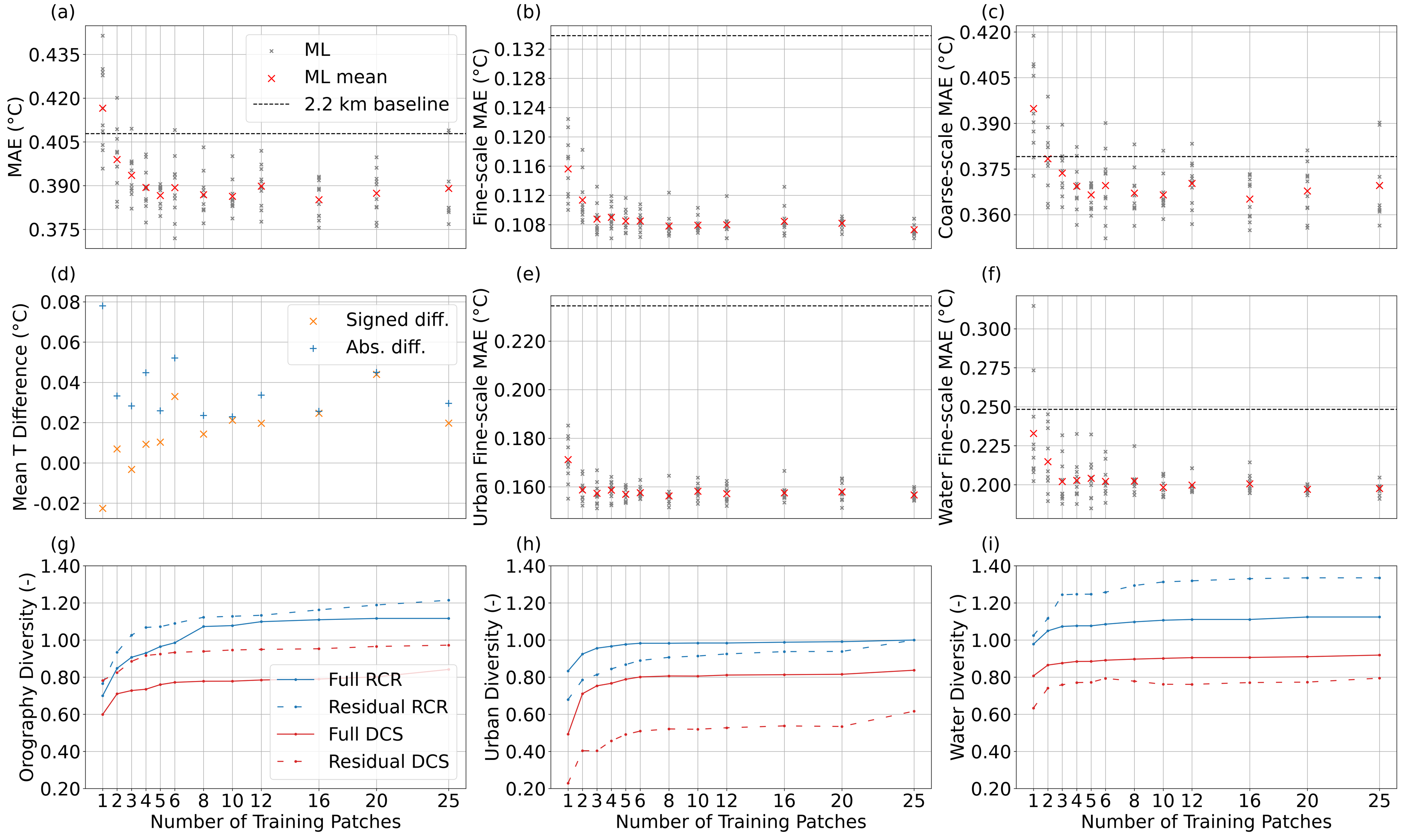}}
  \caption{Test set performance metrics and surface descriptor coverage as a function of the number of training patches. (a) Overall MAE ($^\circ$C), (b) MAE for fine-scale ($<$2.2~km; FS) decomposition, (c) MAE for coarse-scale ($>$2.2~km; CS) decomposition, (d) a single signed mean difference (orange; $\times$) and a single mean absolute difference (blue; $+$) at each training-patch count, computed by averaging the signed and absolute realization-specific offsets from the 10 randomized training realizations after each prediction field is averaged over the full held-out test set; values are compared with the 300~m reference field, (e) MAE restricted to grid points with urban heterogeneity, (f) MAE restricted to grid points with inland water heterogeneity, and (g--i) Range Coverage Ratio (RCR; blue) and Distribution Coverage Score (DCS; red) for full (solid) and residual (dashed) fields of orography (g), urban fraction (h), and inland water fraction (i). In (a--f), horizontal dashed black lines represent the 2.2~km baseline error. Individual gray marks indicate results for each of the 10 randomized realizations, while red marks denote their means.}\label{fig:exp1_scaling}
\end{figure}

\begin{figure}[t]
  \centerline{\includegraphics[width=0.85\textwidth]{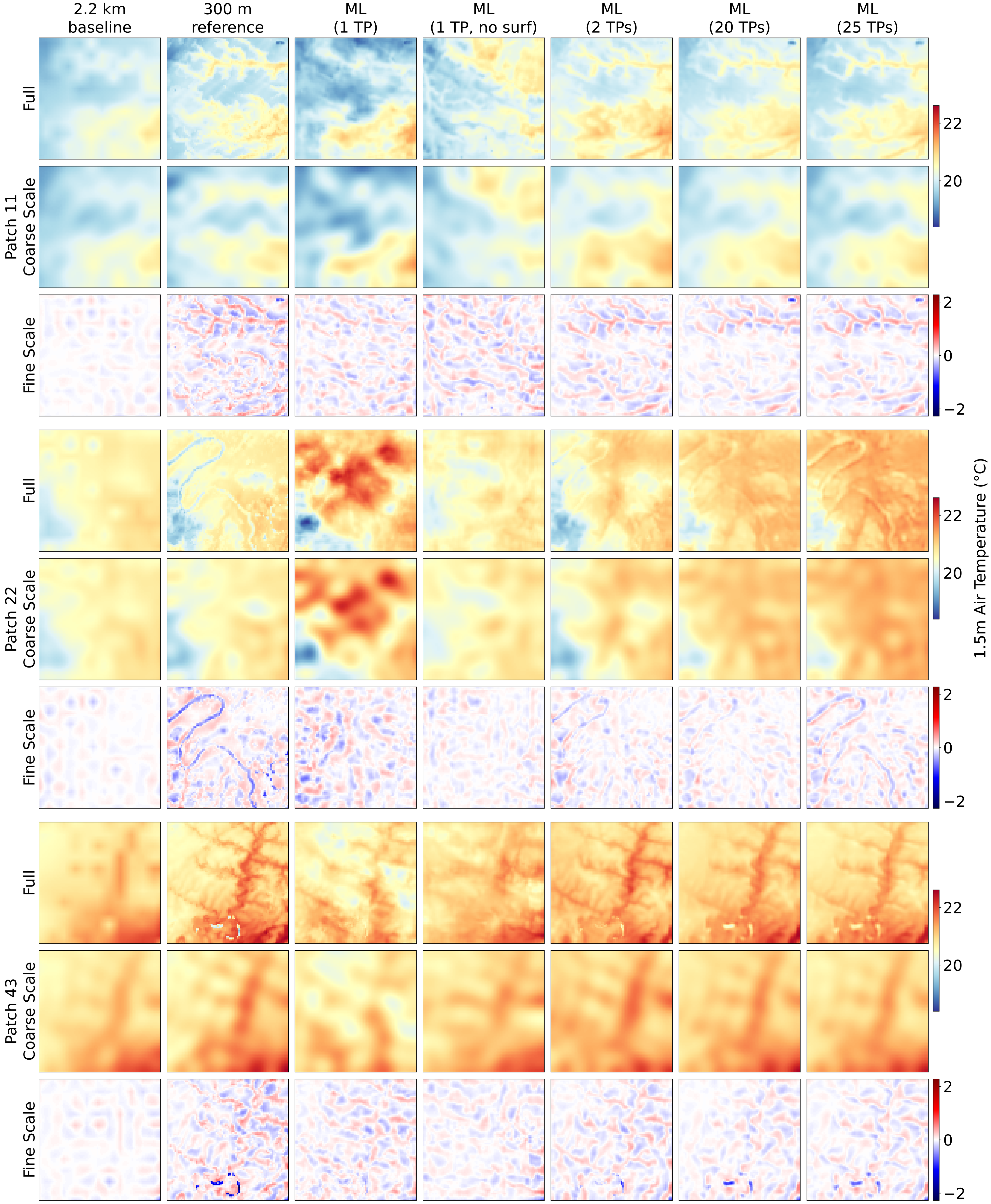}}
  \caption{Temperature predictions for patches 11, 22 and 43 at 09:00 UTC on 2024-08-31. Columns: 2.2~km baseline, 300~m reference, ML trained on patch 12, ML trained on patch 12 without surface variables, ML trained on patches 12 and 34, ML trained on all 20 training patches, and ML trained on the training and test patches. For each patch, there are three rows, consisting of the full, coarse-scale, and fine-scale fields.}\label{fig:exp1_spatial_case}
\end{figure}

\subsubsection{Overall Error Trajectories and Surface Coverage (Exp. 1)}\label{sec:exp1_overall}

Fig.~\ref{fig:exp1_scaling}a illustrates the overall test set MAE as active training patches increase from 1 to 20. When trained on a single spatial patch, the model exhibits an average test set MAE of 0.417~$^\circ$C, slightly underperforming the 2.2~km baseline (0.408~$^\circ$C). Expanding the training set to two patches reduces overall error below the baseline to 0.399~$^\circ$C, a gain visible in spatial maps (Fig.~\ref{fig:exp1_spatial_case}) through resolved cool parks and warm low-lying topography.

A major proportion of the total achievable performance improvement is reached using three training patches (MAE of 0.394~$^\circ$C; a 3.5\% error reduction relative to the baseline). Beyond three patches, performance plateaus at approximately five patches with a minimum out-of-sample MAE of 0.387~$^\circ$C (a 5.6\% improvement). Including test patches directly in training yields no further improvement, indicating that the model learns generalizable relationships.

To examine whether performance saturation is associated with surface descriptor similarity between training and test patches, we evaluate RCR and DCS across expanding training configurations (Fig.~\ref{fig:exp1_scaling}g--i). Range Coverage Ratio (RCR) quantifies how much of the test-set value range is spanned by the training patches, and Distribution Coverage Score (DCS) measures how closely the training and test distributions match. We compute these metrics for orography, urban fraction, and inland water fraction to assess how completely the training data sample the surface-condition space the network must generalize across. As shown in Fig.~\ref{fig:exp1_scaling}g, RCR and DCS for orography plateau as additional patches are added. However, Pearson correlation analysis (not shown) reveals very weak statistical links between these coverage metrics and the MAE across the 10 random realizations. This discrepancy indicates that RCR and DCS do not fully capture the physical patterns required by the network to map 2.2~km inputs to 300~m references, and that the multi-variable use of orography, urban fraction, and inland water fraction is not adequately represented by single-variable distribution metrics. It is possible that the area of applicability method developed by \citet{meyer2021predicting} could better explain the relationship between surface descriptor coverage and model performance, but requires an expensive feature performance analysis, which is beyond the scope of the current study.

\subsubsection{Assessing Ability to Add Spatial Detail (Exp. 1)}\label{sec:exp1_freq}

To assess the ability of ML downscaling to add spatial detail, we apply FFT high-pass filtering using a 2.2~km cutoff scale. This isolates the sub-2.2~km temperature detail added by the ML model beyond what the 2.2~km driving model can resolve. We compute the MAE between the fine-scale components of the ML predictions and the 300~m model reference. For comparison, this evaluation is repeated for the 2.2~km baseline model. The 2.2~km model serves as both a source of predictor variables for the ML model and as a performance baseline. Because the 2.2~km baseline contains virtually no signal at sub-2.2~km scales, its fine-scale error is approximately equal to the magnitude of the 300~m reference fine scale itself.

Fine-scale spatial skill emerges rapidly under minimal training patches. As shown in Fig.~\ref{fig:exp1_scaling}b, a model trained on a single spatial patch achieves a fine-scale MAE 13.6\% lower than the baseline. This demonstrates that the network utilizes local surface descriptors to synthesize sub-grid structures that generalize to unobserved regions, capturing fine-scale thermal gradients along river valleys and ridges (Fig.~\ref{fig:exp1_spatial_case}). Fine-scale MAE improvements plateau by three training patches at approximately 19.3\% over the baseline.

The coarse-scale MAE is the mean absolute error between the prediction (ML and 2.2~km baseline) and the 300~m reference at scales coarser than 2.2~km. The ML is guided by the 2.2~km model since it is an input, but in training it is free to learn its own mapping between the 2.2~km model and 300~m model across spatial scales. Thus, the coarse-scale MAE measures how well the downscaled field reproduces broad temperature structure and systematic differences between the 2.2~km driving field and the 300~m reference. Discrepancies between the coarse-scale 300~m reference and the 2.2~km baseline arise because unresolved fine-scale processes (e.g., turbulent mixing or microclimatic surface interactions) systematically warm or cool the broader temperature state. The coarse-scale MAE behaves similarly to the fine-scale component, improving gradually but plateauing slightly later than the fine-scale component at roughly five training patches, with a 3.3\% error reduction relative to the 2.2~km baseline (Fig.~\ref{fig:exp1_scaling}c). While the relative improvement at coarse scales (3.3\%) is smaller than at fine scales (19.3\%), the higher baseline error at coarse scales means that absolute MAE reductions are more comparable (0.015~$^\circ$C for coarse scales versus 0.027~$^\circ$C for fine scales). Nonetheless, the percentage reduction confirms that the primary value added by the ML model remains at the fine scales.

Single-patch training exhibits elevated coarse-scale error compared to the baseline due to spatial overfitting, where localized thermal offsets from the training set are incorrectly applied to the test set. This is confirmed in Fig.~\ref{fig:exp1_scaling}d, where the signed mean difference is the average of the signed realization-specific offsets and the mean absolute difference is the average of the absolute realization-specific offsets. The signed mean difference is small because some realizations are biased warm and others cold, so their offsets partially cancel in the mean, while the mean absolute difference still captures the typical magnitude of the offset, which is large. Here, the 300~m reference is the held-out high-resolution target field. Expanding training to two patches drops the mean absolute difference sharply from 0.078~$^\circ$C to 0.033~$^\circ$C (Fig.~\ref{fig:exp1_scaling}d), alongside a modest (0.016~$^\circ$C) reduction in the signed mean difference. This demonstrates that multi-patch training mitigates spatial overfitting, reducing both systematic bias and realization-specific thermal offsets.

\subsubsection{Performance over Heterogeneous Surface Features (Exp. 1)}\label{sec:exp1_hetero}

Much of the added value of 300~m simulations arises from their ability to represent temperature differences over heterogeneous local surface conditions, such as urban-rural and land-water contrasts. Here we evaluate ML downscaling from 2.2~km to 300~m over localized heterogeneous surface features using two separate masks: one based on urban fraction and one based on inland water fraction. In each case, grid cells are flagged as heterogeneous when the corresponding fraction differs by more than $0.1$ from at least one neighboring grid cell.

MAE over urban heterogeneous grid points plateaus after exposure to two training patches (Fig.~\ref{fig:exp1_scaling}e). Over inland water boundaries (Fig.~\ref{fig:exp1_scaling}f), MAE stabilizes at three training patches, showing subtle continued gains as training diversity increases. This delayed stabilization likely relates to the extreme spatial sparsity of water features within the training set; because few grid cells contain inland water (Fig.~\ref{f1}c), broader spatial sampling provides necessary exposure to water--land thermal contrasts. RCR and DCS metrics do not strongly reflect this specific sample-volume dependency over water features.

\subsubsection{Role of Static Surface Predictors and Spatial Overfitting (Exp. 1)}\label{sec:exp1_no_surface}

We have shown that the ML downscaling has skill in spatially transferring relationships learned in other patches. To test the importance of the surface descriptors, we repeat the transferability test using a single training patch (patch 12) while withholding all high-resolution static surface and orographic predictors.

Omitting static surface descriptors severely degrades spatial transferability (Fig.~\ref{fig:exp1_spatial_case}, Column 4 vs. Column 3). Without surface inputs, the network projects a persistent imprint of the river Seine and topographical structures from patch 12 (Fig.~\ref{f1}a) directly onto the test set. Rather than learning physical relationships between surface properties and thermal responses, the network exhibits severe spatial overfitting tied to grid-coordinate memory.

Here we have shown that incorporating the high-resolution static surface descriptors was required to obtain the spatial transferability demonstrated above, preventing spatial overfitting from degrading performance when transferring to unseen locations.

\subsection{Experiment 2: Transferability Across Orographic-Complexity Shift}\label{sec:exp2_results}

In this experiment, we investigate how well relationships learned over regions of different terrain complexity transfer to locations with more complex topography.

The baseline model is trained on three low-complexity patches (patches 41, 00, and 31; $99^{\text{th}}$ percentile elevation gradients of 0.035, 0.041, and 0.048~$\text{m~m}^{-1}$) and evaluated on three complex test patches (patches 44, 11, and 02; $99^{\text{th}}$ percentile elevation gradients of 0.107, 0.108, and 0.109~$\text{m~m}^{-1}$). To test the impact of introducing higher-slope features during training, one low-complexity patch (31) is progressively replaced: first with a medium-complexity patch (patch 32; $99^{\text{th}}$ percentile gradient of 0.08~$\text{m~m}^{-1}$, denoted ``Low+Medium''), and subsequently with the most complex patch available (patch 03; $99^{\text{th}}$ percentile gradient of 0.14~$\text{m~m}^{-1}$, denoted ``Low+Highest''). All models are evaluated on the same three complex test locations. To isolate the role of residual orography (300~m minus 2.2~km baseline) in facilitating transferability to unseen slope regimes, this experiment is repeated with the residual orography predictor omitted.

\begin{figure}[ht]
  \centerline{\includegraphics[width=0.95\textwidth]{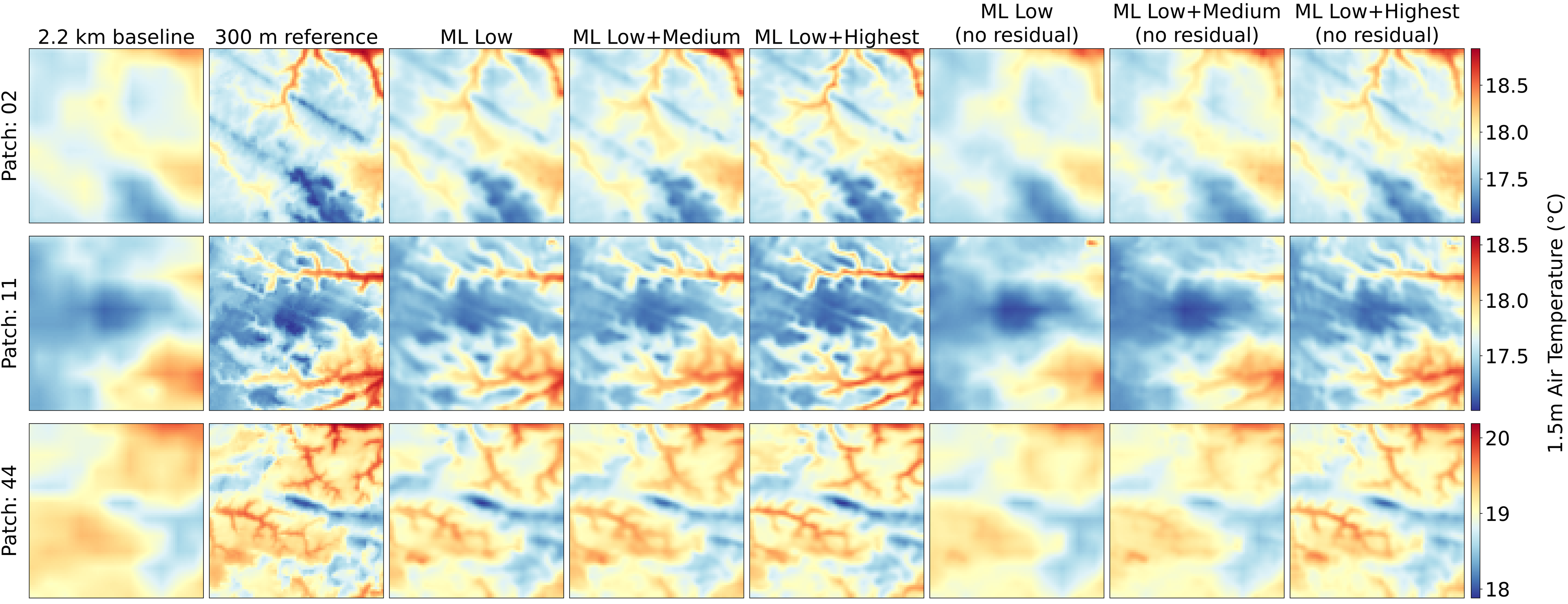}}
  \caption{Temperature predictions for complex test patches 02, 11, and 44 at 07:00 UTC on 2024-09-02 across varying training set terrain complexities and predictor configurations. Columns from left to right: 2.2~km baseline, 300~m reference, ML (Low), ML (Low+Medium), ML (Low+Highest), ML (Low, no residual), ML (Low+Medium, no residual), and ML (Low+Highest, no residual).}\label{f4}
\end{figure}

\begin{table}[ht]
\caption{Overall, coarse-scale ($>$2.2~km), and fine-scale ($<$2.2~km) Mean Absolute Error (MAE; $^\circ$C) evaluated across full test patches (02, 11, 44) and restricted to masked steep-gradient locations beyond the training slope range across different training set terrain complexities, with and without the inclusion of the residual orography predictor. Reported MAEs are means over the three training realizations for each configuration.}
\label{t1}
\centering
\setlength{\tabcolsep}{5pt}
\resizebox{\textwidth}{!}{
\begin{tabular}{lccccccc}
\hline
\textbf{Metric} & \textbf{2.2~km Baseline} & \textbf{Low} & \textbf{Low+Medium} & \textbf{Low+Highest} & \textbf{Low (No Resid.)} & \textbf{Low+Medium (No Resid.)} & \textbf{Low+Highest (No Resid.)} \\
\hline
Overall MAE & 0.414 & 0.395 & 0.399 & 0.403 & 0.409 & 0.413 & 0.403 \\
Coarse-Scale MAE & 0.381 & 0.369 & 0.374 & 0.378 & 0.379 & 0.386 & 0.377 \\
Fine-Scale MAE & 0.149 & 0.126 & 0.124 & 0.123 & 0.140 & 0.132 & 0.126 \\
\hline
Overall MAE (Masked) & 0.381 & 0.371 & 0.374 & 0.376 & 0.380 & 0.387 & 0.379 \\
Coarse-Scale MAE (Masked) & 0.357 & 0.352 & 0.356 & 0.360 & 0.357 & 0.366 & 0.359 \\
Fine-Scale MAE (Masked) & 0.131 & 0.109 & 0.108 & 0.105 & 0.122 & 0.116 & 0.110 \\
\hline
\end{tabular}
}
\end{table}

\subsubsection{Impact of Training Complexity on Scale-Dependent Errors (Exp. 2)}\label{sec:exp2_complexity_errors}

As summarized in Table~\ref{t1}, training on low complexity yields a lower overall MAE (0.395~$^\circ$C) than the 2.2~km baseline (0.414~$^\circ$C; a 4.6\% improvement). Including patches with higher elevation gradients slightly increases overall MAE to 0.399~$^\circ$C (``Low+Medium'') and 0.403~$^\circ$C (``Low+Highest''). This increase is concentrated in the coarse-scale component, which increases from 0.369~$^\circ$C to 0.378~$^\circ$C. Thus, exposing the model to more complex terrain does not necessarily improve its representation of the broader temperature field. It is unlikely that this slight coarse-scale degradation represents a systematic penalty of complex terrain; rather, as shown in Fig.~\ref{fig:exp1_scaling}c, coarse-scale errors exhibit substantial realization-to-realization variability depending on the specific combination of training patches selected.

Conversely, fine-scale ($<$2.2~km) spatial skill systematically improves as terrain complexity is added to training. Relative to the baseline (0.149~$^\circ$C), fine-scale MAE improves from 0.126~$^\circ$C (``Low'') to 0.124~$^\circ$C (``Low+Medium'') and 0.123~$^\circ$C (``Low+Highest''). Introducing complex terrain enables the model to better resolve localized topographically driven temperature patterns, capturing the contrasting thermal responses of valleys and elevated terrain with greater fidelity (Fig.~\ref{f4}).

To isolate model behavior at steep slope features, metrics were evaluated over a masked subset of test grid points where elevation gradients exceed the maximum bounds sampled during training (Table~\ref{t1}). These steep-slope points beyond the training range account for 4.1\% and 1.0\% of the test set for ``Low'' and ``Low+Medium'' configurations, respectively. On these masked points, fine-scale MAE improves from 0.109~$^\circ$C (``Low'') to 0.108~$^\circ$C (``Low+Medium'') and reaches 0.105~$^\circ$C when all points fall within the training slope range. These differences are small, so we do not interpret the incremental improvements between training configurations as evidence of a statistically significant effect. Instead, the results show that the model retains strong fine-scale skill over slope values that were absent from the training data, while additional exposure to complex terrain provides only modest further improvement.

\subsubsection{Role of Residual Orography as a Predictor (Exp. 2)}\label{sec:exp2_residual}

It has been shown that transfer of training on 300~m patches improves errors in predictions, with a suggestion of a small role for choices of orographic complexity in choices of training patches. However, omitting residual orography degrades performance across all training configurations (Table~\ref{t1}), causing a visible loss of fine-scale detail along steep terrain boundaries (Fig.~\ref{f4}). The reliance on residual orography depends on the terrain complexity of the training set. When trained on simple terrain (``Low''), omitting residual orography increases fine-scale MAE by 10.7\% overall (0.126~$^\circ$C to 0.140~$^\circ$C) and by 11.9\% over masked steep points (0.109~$^\circ$C to 0.122~$^\circ$C). As training complexity increases, this penalty systematically declines to 6.2\% (full) / 7.4\% (masked) for ``Low+Medium'' and 2.4\% (full) / 4.8\% (masked) for ``Low+Highest''.

Thus, residual orography acts as a critical structural aid when training data lacks physical topographic diversity. By explicitly encoding resolution discrepancies, the residual field highlights sharp gradients, enabling the network to downscale effectively even from gentle terrain. As complex topographic features are introduced directly via static surface descriptors, reliance on the residual field diminishes.

Points exceeding the training bounds in residual gradient space were found to be strongly collocated with points exceeding the training bounds in absolute gradient space. The primary utility of residual orography is therefore not altering distribution coverage bounds; rather, it likely enhances fine-scale topographic signals during training on subdued terrain.

\subsection{Experiment 3: Transferability Across Elevation Shift}\label{sec:exp3_results} 

Elevation has an important impact on temperature. Here, we test whether relationships learned during training can be transferred to predict temperatures in regions whose elevations fall outside the training range.

In this final experiment, we investigate transferability across distinct altitude regimes, evaluating generalization from low-lying to elevated topography under a train--test elevation shift. The baseline model is trained on three low-elevation patches (patches 22, 34, and 21; mean altitudes of 62.1~m, 78.2~m, and 82.0~m) and evaluated on three high-elevation test patches (patches 10, 11, and 00; mean altitudes of 138.8~m, 143.5~m, and 148.0~m). To evaluate the impact of exposing the model to high-elevation regimes during optimization, one low-lying training patch (21) is replaced with the highest-elevation patch in the domain (patch 01; mean altitude of 151.2~m, denoted ``Low+Highest'') and tested across the same three elevated test locations. This experiment is repeated with the residual orography predictor removed to evaluate whether it assists the network in learning lapse-rate-driven thermal changes across altitude regimes.

\begin{figure}[ht]
  \centerline{\includegraphics[width=0.9\textwidth]{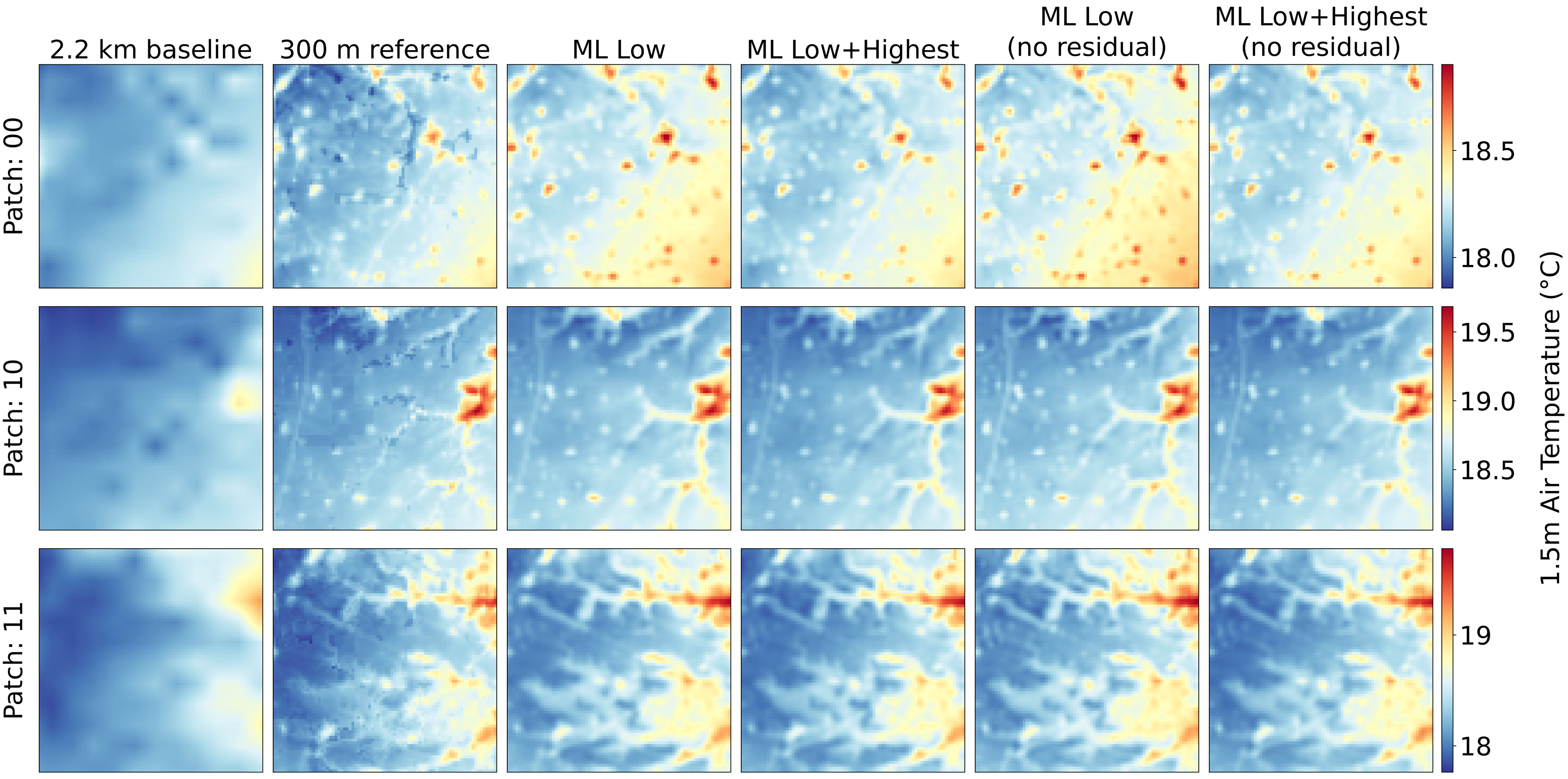}}
  \caption{Time-averaged temperature fields across elevated test patches 00, 10, and 11 for different training set altitude configurations and predictor choices. Columns from left to right: 2.2~km baseline, 300~m reference, ML (Low), ML (Low+Highest), ML (Low, no residual), and ML (Low+Highest, no residual).}\label{f5}
\end{figure}

\begin{table}[ht]
\caption{Overall, coarse-scale ($>$2.2~km), and fine-scale ($<$2.2~km) Mean Absolute Error (MAE; $^\circ$C) evaluated across elevated test patches (10, 11, 00) for low and low+highest training altitude regimes, with and without the inclusion of the residual orography predictor. Reported MAEs are means over the three training realizations for each configuration.}
\label{t2}
\centering
\setlength{\tabcolsep}{6pt}
\begin{tabular}{lccccc}
\hline
\textbf{Metric} & \textbf{2.2~km Baseline} & \textbf{Low} & \textbf{Low+Highest} & \textbf{Low (No Resid.)} & \textbf{Low+Highest (No Resid.)} \\
\hline
Overall MAE ($^\circ$C) & 0.383 & 0.380 & 0.373 & 0.394 & 0.391 \\
Coarse-Scale MAE ($^\circ$C) & 0.362 & 0.364 & 0.357 & 0.378 & 0.375 \\
Fine-Scale MAE ($^\circ$C) & 0.106 & 0.090 & 0.089 & 0.092 & 0.092 \\
\hline
\end{tabular}
\end{table}

\subsubsection{Impact of Training Altitude on Broad Thermal Biases (Exp. 3)}\label{sec:exp3_altitude_errors}

As shown in Table~\ref{t2}, training exclusively on low-altitude locations yields an overall test set MAE of 0.380~$^\circ$C across the elevated test patches (a 0.8\% improvement realtive to the 2.2~km baseline (0.383~$^\circ$C)). Incorporating the highest-elevation patch during training (``Low+Highest'') lowers overall MAE to 0.373~$^\circ$C (a 2.6\% improvement relative to the 2.2~km baseline).

This improvement is driven almost entirely by the coarse-scale component, where MAE decreases from 0.364~$^\circ$C for ``Low'' to 0.357~$^\circ$C for ``Low+Highest''. When trained purely on low-lying terrain, the model fails to capture the cooler mean temperatures characteristic of higher elevations, yielding a coarse-scale MAE (0.364~$^\circ$C) that slightly exceeds the 2.2~km baseline (0.362~$^\circ$C). Because lapse-rate cooling causes elevated regions to be cooler on average, transferring a low-altitude model to higher elevations induces a persistent warm bias (Fig.~\ref{f5}, patch 00). Adding the highest-elevation patch to training reduces this elevation-dependent mean bias, reducing coarse-scale MAE to 0.357~$^\circ$C.

In contrast, fine-scale ($<$2.2~km) spatial errors remain nearly identical, shifting from 0.090~$^\circ$C (``Low'') to 0.089~$^\circ$C (``Low+Highest''). Shifting the mean altitude of a training patch alters the broad thermal state but does not introduce new fine-scale terrain complexity or steeper gradients. Fine-scale spatial structures at 300~m resolution are thus already well-captured by the low-altitude model, while higher-elevation data primarily constrains the broader-scale thermal adjustment.

\subsubsection{Role of Residual Orography in Learning Elevation-Dependent Biases (Exp. 3)}\label{sec:exp3_residual_elevation}

Residual orography is defined as the 300~m orography minus the 2.2~km orography bilinearly interpolated onto the 300~m grid, so it captures the fine-scale topographic structure that is unresolved by the 2.2 km model. Here, we investigate its role in improving transferability across elevation regimes and in reducing patch-scale thermal biases under elevation-shift conditions.

Omitting the residual orography predictor degrades performance across all elevation regimes (Table~\ref{t2}). For ``Low'' and ``Low+Highest'' setups, removing residual orography increases overall MAE from 0.380~$^\circ$C to 0.394~$^\circ$C and from 0.373~$^\circ$C to 0.391~$^\circ$C, respectively, causing ML to perform worse than the 2.2~km baseline (0.383~$^\circ$C).

This degradation is concentrated in the coarse-scale component, where MAE increases from 0.364~$^\circ$C to 0.378~$^\circ$C for ``Low'' and from 0.357~$^\circ$C to 0.375~$^\circ$C for ``Low+Highest''. By contrast, fine-scale MAE increases only marginally from 0.090~$^\circ$C to 0.092~$^\circ$C for ``Low'' and from 0.089~$^\circ$C to 0.092~$^\circ$C for ``Low+Highest''.

The time-averaged warm bias observed in the ``Low'' model persists regardless of whether residual orography is included (Fig.~\ref{f5}, patch 00). Thus, although residual orography improves the overall coarse-scale error, it does not remove the systematic warm bias associated with applying a model trained only at low elevations to the higher-elevation test patches.

Inspection of individual timesteps indicates that coarse-scale errors at patch scale vary dynamically over time, fluctuating non-systematically between over- and under-estimation across different forecast states. Including residual orography reduces these transient coarse-scale errors, yielding a lower overall coarse-scale MAE. However, because the persistent warm bias remains, the results do not establish that residual orography is providing a lapse-rate correction. The precise mechanism by which sub-grid topographic residuals influence these time-varying patch-scale biases remains unexplained.

\section{Conclusions}\label{sec:conclusions}

In this study we demonstrate the capability to enhance temperature resolution to 300~m over the Paris region using relationships spatially transferred from elsewhere in the domain, and use controlled experiments to investigate the components that enable this generalization. We systematically investigate zero-shot spatial transferability of a machine learning downscaling (MLD) model for temperature within a geographically and climatologically similar domain, emulating 300~m ensemble simulations over the Paris region. By isolating the effects of surface heterogeneity and orography from confounding climatic shifts, the findings provide clear insights into the spatial generalization capabilities and limits of deep learning downscaling architectures.

The key conclusions of this work are summarized as follows:
\begin{enumerate}
    \item \textbf{Spatial Training Diversity and Saturation:} Fine-scale spatial details emerge rapidly under minimal spatial training exposure. A model trained on a single spatial patch partially represents localized thermal contrasts -- including fine-scale features across complex urban areas, land-water interfaces, cool parks, and low-lying orography -- reducing fine-scale MAE against the 300~m reference by 13.6\% relative to the 2.2~km baseline. However, a single patch yields elevated coarse-scale errors due to patch-dependent mean bias overfitting. Expanding the training set to two to three patches prevents the network from mapping patch-specific mean temperature offsets as generalizable spatial signals, reducing the aggregated mean absolute offset across the 10 randomized training realizations from 0.078~$^\circ$C to 0.033~$^\circ$C. Overall MAE plateaus after exposure to approximately five patches (achieving a 5.6\% error reduction over the baseline), with fine-scale errors stabilizing at three patches and coarse-scale errors at five patches. Crucially, including held-out test patches directly in training yields no further performance improvement, confirming that the model learns physically generalizable relationships rather than memorizing location-specific features.

    \item \textbf{Role of Static Surface Descriptors and Distributional Metrics:} Incorporating high-resolution static surface descriptors is crucial for spatial transferability. Omitting surface variables leads to severe spatial overfitting, which with one training patch leads to persistent imprints of localized training features (such as river channels) projecting directly onto unseen test patches. While surface coverage metrics -- Range Coverage Ratio and Distribution Coverage Score -- plateau alongside MAE trajectories, they exhibit very weak statistical links with test set MAE across random realizations. This indicates that simple univariate range or distribution metrics do not fully capture the complex, multi-variate feature combinations or geometric contexts required for neural networks to learn accurate atmospheric-to-surface mappings.

    \item \textbf{Transferability Across Orographic-Complexity Shift:} When trained exclusively on low-complexity terrain and residual orography is included as a predictor, the model retains strong transferability to complex, high-gradient slope regions, outperforming the 2.2~km baseline (overall MAE of 0.395~$^\circ$C vs. 0.414~$^\circ$C baseline; a 4.6\% improvement). Introducing topographically complex patches into training yields only incremental fine-scale spatial gains -- reducing fine-scale MAE from 0.126~$^\circ$C to 0.123~$^\circ$C, and masked steep-slope MAE from 0.109~$^\circ$C to 0.105~$^\circ$C. This supports the interpretation that the model is able to learn slope-related temperature dependence from the orography predictors, and that transfer from simpler to more complex terrain remains broadly robust so long as residual orography is included.

    \item \textbf{Transferability Across Elevation Shift Regimes:} Training strictly on low-lying terrain introduces a patch-scale warm bias when evaluating over elevated topography, as the network fails to capture lapse-rate cooling. This degradation is confined to the coarse-scale component (coarse-scale MAE of 0.364~$^\circ$C vs. 0.362~$^\circ$C baseline). Introducing high-altitude patches during training reduces this mean thermal bias, bringing coarse-scale MAE down to 0.357~$^\circ$C and overall MAE to 0.373~$^\circ$C (a 2.6\% improvement over the baseline), while fine-scale skill remains largely unaffected (0.090~$^\circ$C vs. 0.089~$^\circ$C).

    \item \textbf{Structural Value of Residual Orography:} Ablation tests show that the residual surface descriptor field ($300\text{~m} - 2.2\text{~km}$) is a key enabler of robust terrain transfer. Omitting residual orography severely degrades spatial transferability across all terrain types -- increasing fine-scale MAE by 10.7\% over simple terrain (and by 11.9\% over steep slope points) and driving overall MAE above the 2.2~km baseline when evaluating across elevation regimes (0.394~$^\circ$C vs. 0.383~$^\circ$C baseline). The residual field sharpens fine-scale topographic signals when training topographic complexity is low. Although including residual orography reduces time-varying coarse-scale errors across elevation regimes, inspection of individual timesteps reveals that these patch-scale biases fluctuate dynamically over time, leaving the exact mechanism behind this stabilization unexplained.
\end{enumerate}

These findings demonstrate that MLD offers a viable, computationally efficient mechanism to emulate hectometre-scale NWP simulations, providing a scalable strategy to extend urban-scale forecasting from city-scale coverage to regional and continental scales.

The results also show that spatial transferability challenges persist across different physical scales. At fine spatial scales ($<$2.2~km), thermal responses to static surface features generalize robustly without degrading performance below the driving model. In contrast, large-scale ($>$2.2~km) altitude-dependent temperature differences between the 2.2~km and 300~m models can cause MLD to underperform the 2.2~km model under large train-test altitude shifts. This may be linked to the use of independent 2.2~km and 300~m model fields: because the MLD maps directly from the 2.2~km driver to the 300~m reference, rather than from coarse-grained 300~m fields, the network must reconcile substantial large-scale forecast differences, potentially making it more susceptible to coarse-scale errors. Using coarse-grained 300~m fields as inputs could constrain the large-scale predictions more closely to the driving field, potentially limiting spurious coarse-scale deviations across unseen altitude regimes and forecast lead times, but would not provide the same opportunity to correct systematic differences between the 2.2~km driver and 300~m reference.

\clearpage

\acknowledgments
All authors were supported by the AI4Climate Project  programme, funded by the UK Government’s Department for Science, Innovation and Technology (DSIT) through the International Science Partnerships Fund (ISPF).

\datastatement
Data and code used in this study is proprietary and cannot be shared publicly. 

\bibliographystyle{ametsocV6}
\bibliography{references}

\end{document}